\documentstyle[aps,prb,multicol,floats,epsf]{revtex}
\begin{document}
\draft
\preprint{}
\wideabs{

\title{
Temperature dependence of spin correlation and charge dynamics\\
in the stripe phase of high-$T_c$ superconductors
}

\author{Y. Shibata, T. Tohyama, and S. Maekawa}
\address{Institute for Materials Research, Tohoku University,
        Sendai 980-8577, Japan}
\date{Received 10 November 2000; revised manuscript recieved 15 February 2001}
\maketitle

\begin{abstract}
We examine the temperature dependence of the electronic states
in the stripe phase of high-$T_c$ cuprates by using the $t$-$J$
model with a potential that stabilizes vertical charge stripes.
Charge and spin-correlation functions and optical conductivity
are calculated by using finite-temperature Lanczos method.
At zero temperature, the antiferromagnetic correlation between
a spin in a charge stripe and that in a spin domain adjacent
to the stripe is weak, since the charge stripe and the spin
domain are almost separated.  With increasing temperature,
the correlation increases and then decreases toward high
temperature.  This is in contrast to other correlations that
decrease monotonically.  From the examination of the charge
dynamics, we find that this anomalous temperature dependence
of the correlation is the consequence of a crossover from
one-dimensional electronic states to two-dimensional ones.
\end{abstract}
}
\narrowtext

\section{Introduction}
Charge stripes and related phenomena are now hot topics in the
high-$T_c$ superconductor research field. Neutron-scattering
measurements performed on La$_{1.475}$Nd$_{0.4}$ Sr$_{0.125}$CuO$_4$
(LNSC) with the low-temperature tetragonal (LTT) structure have
revealed the presence of a charge order and an incommensurate
magnetic order.~\cite{Tran} These charge and spin orders have
been explained by assuming a stripe structure that consists of
vertical charge stripes and antiphase between spin domains.
The incommensurate magnetic order has been observed not only
in LNSC but also in La$_{2-x}$Sr$_x$CuO$_4$ (LSC) with
$x\simeq0.12$ carrier doping.~\cite{Yamada,Suzuki,Kimura}
Around this carrier concentration, an incomplete phase transition
from the low-temperature orthorhombic (LTO) phase to the LTT
phase has been observed.~\cite{Goto,Moo,Sai,Boz}  While clear
evidence of the charge order has not been reported, it is
considered that the electronic states in LSC are similar to
those in the LNSC.

Anomalous behaviors that are supposed to be related to the
presence of the stripes have been observed in the angle-resolved
photoemission (ARPES) spectrum as suppressed spectral weight
along the (0,0)-($\pi$,$\pi$) direction~\cite{Ino} and in
the optical conductivity as the enhancement of intensity in
the mid-infrared region.~\cite{Tajima}  These features have
been explained by the present authors, by using the exact
diagonalization calculation at zero-temperature for a model
that includes both the strong electron correlation and the stripes,
i.e., a $t$-$J$ model with an additional potential introduced to
stabilize vertical charge stripes.~\cite{Tohyama}  The ground
state of the model is characterized by the charge stripes
along which the charge carriers can move coherently, but
perpendicular to which charge carriers show only incoherent
motion.  This situation is consistent with the behavior of the
Hall coefficient at low temperature~\cite{Noda} as well as
the ARPES data~\cite{Zhou} in LNSC, both of which indicate
the one-dimensional motion of the charge carriers. At the same
time, the zero temperature calculation~\cite{Tohyama} has shown
that the spin correlation inside the spin domains is as strong
as those for the Heisenberg spin system, whereas the correlation
between a spin in a spin domain and that in a charge stripe
adjacent to the domain is weak since the charge stripe and
the spin domain are almost separated.

With the increase of temperature, the one-dimensional electronic
states in the stripe phase mentioned above are expected to be
destroyed and evolved into the two-dimensional ones.  Therefore
it is interesting to know how such evolution occurs and how
the physical quantities are affected by the evolution.
In this study, we clarify the temperature dependence of the
electronic state in the stripe phase.  By using the
finite-temperature Lanczos method developed by Jakli\v{c} and
P. Prelov\v{s}ek,~\cite{Jaklic1,Jaklic2} we calculate the optical
conductivity and spin-correlation functions at finite temperatures
in the $t$-$J$ model with a potential that stabilizes the vertical
charge stripes.  We find that the spin correlation between a spin
in the charge stripe and that in the spin domain increases with
increasing temperature and then decreases toward higher
temperature.  This is in contrast to other spin correlations
that decrease monotonically.  By examining the dependence of
charge dynamics on temperature, this anomalous temperature
dependence of the spin correlation is concluded to be
the consequence of a crossover from one-dimensional electronic
states to two-dimensional ones.

We introduce our model, i.e., a $t$-$J$ model with a stripe
potential, and show outlines of the finite temperature Lanczos
method in Sec.~II.  In Sec.~III, results and discussions on the
temperature dependence of spin-and charge-correlation functions
and the optical conductivity are presented.  A possible method
to confirm experimentally the evolution of the electronic states
in stripe phase with temperature is also proposed.
The summary is given in Sec.~IV.

\section{Model and Numerical Method}
We introduce the $t$-$J$ model, which is given by
\begin{eqnarray}
H_{tJ}&=&
    J\sum\limits_{\left\{i,j\right\}} {\bf S}_i\cdot {\bf S}_j
    -t\sum\limits_{\left\{i,j\right\} \sigma }
    (\tilde{c}_{i\sigma}^\dagger \tilde{c}_{j\sigma}+{\rm H.c.})\;,
\label{HtJ}
\end{eqnarray}
where $\tilde{c}_{i\sigma}=c_{i\sigma}(1-n_{i-\sigma})$ is the
annihilation operator of an electron with spin $\sigma$ at site $i$
with the constraint of no double occupancy, ${\bf S}_i$ is the spin
operator, and the summation $\left\{i,j\right\}$ runs over the
nearest-neighbor pairs.

It is controversial whether the $t$-$J$ model itself has
the stripe-type ground state.~\cite{White,Hell,Koba,Toh}
A possible origin of the appearance of stable stripe phase is
due to the presence of the long-range part of the Coulomb
interaction~\cite{Emery,Sei,Vei} and/or the coupling to lattice
distortions.  In LSC, the LTT fluctuation seems to help the
latter mechanism.~\cite{Sai,Boz}  In fact, the LTT structure
makes Cu-O bonds anisotropic, leading to directional distribution
of carriers through the anisotropy of the Madelung potential
at in-plane oxygen sites and that of the hopping amplitude
between Cu and O.

To model the directional charge distribution and the tendency
toward the stripe instability as simply as possible, we introduce
a configuration-dependent stripe potential $V_S(n_h)$.~\cite{Tohyama}
The magnitude of the stripe potential is assumed to depend on the
number of holes $n_h$ in each column:  $V_S(n_h)=0$ for $n_h$=0
and 1 and $V_S(n_h)=-n_hV$ for $n_h\ge$2 with $V>0$.
$V_S(n_h)$ behaves like an attractive potential for holes
independent of distance between holes.  In the following, we use
a 4$\times$4 cluster of the $t$-$J$ model with two holes to simulate
the underdoped system.  Among four columns in the cluster, the
stripe potential $V_S(n_h)$ is introduced into the second column
from the left.  Periodic and open boundary conditions are imposed
along the directions parallel and perpendicular to the column,
respectively.~\cite{AntiPhase}

We set parameters $J/t$=0.4~($t\simeq$ 0.35 eV), and $V/t$=1 to obtain
the ground state with charge stripes.~\cite{Tohyama}  Here, we note
that, if $V=0$, the ground state in small clusters is not the stripe
state but a uniform state as discussed by Hellberg and
Manousakis~\cite{Hell} and the energy difference between the two
states is of the order of $J$.~\cite{Tohyama2}  In order for the
stripe phase to be in the ground state, it is necessary to introduce
$V$ with a magnitude of more than $J$.  Therefore $V/t$=1 is chosen
to make the stripe state stable enough.  Examining the $V$
dependence of the spin and charge correlation functions discussed
in the next section, we have found that their behaviors are not
altered qualitatively if $V>J$.

We employ the finite temperature Lanczos method.~\cite{Jaklic1,Jaklic2}
Outlines of this method are as follows.  The statistical expectation
value of an operator $\hat{A}$ is given by
\begin{equation}
\left<\hat{A}\right>=\frac{1}{Z}\sum\limits^L_{l=1}
\left<\Psi_l\left|e^{-\beta H} \hat{A}\right|\Psi_l\right>\;,
\label{O1}
\end{equation}
where $Z$ is the partition function defined as
$Z=\sum\limits^{L}_{l=1}\left<\Psi_l\left|e^{-\beta H}\right|\Psi_l\right>$,
$\left\{\left|\Psi_l\right>\right\}$ is a complete basis set,
$L$ is the dimension of the Hamiltonian, and $\beta=1/k_{\rm B}T$,
$k_{\rm B}$ being the Boltzmann factor.  Hereafter, $k_{\rm B}$ is
set to be 1.  In order to obtain eigenvectors and eigenvalues,
we use the Lanczos procedure. We set an arbitrary vector
$\left|\Psi_l\right>$ as the initial vector of the Lanczos step
$\left|\psi^l_1\right>$, i.e., $\left|\psi^l_1\right>=\left|\Psi_l\right>$,
and calculate $\left|\psi^l_2\right>, \left|\psi^l_3\right>, \ldots,
\left|\psi^l_M\right>$ step by step,
\begin{eqnarray}
H\left|\psi^l_1\right>&=&a^l_1\left|\psi^l_1\right>+b^l_2\left|\psi^l_2\right>,
\nonumber \\
H\left|\psi^l_2\right>&=&b^l_2\left|\psi^l_1\right>+a^l_2\left|\psi^l_2\right>
+b^l_3\left|\psi^l_3\right>, \nonumber \\
\vdots \nonumber \\
H\left|\psi^l_{M-1}\right>&=&b^l_{M-1}\left|\psi^l_{M-2}\right>
+a^l_{M-1}\left|\psi^l_{M-1}\right>+b^l_M\left|\psi^l_M\right>, \nonumber \\
H\left|\psi^l_M\right>&=&b^l_M\left|\psi^l_{M-1}\right>
+a^l_M\left|\psi^l_M\right>
\;,
\label{Lan}
\end{eqnarray}
where $M$ is a given maximum number of Lanczos steps.  Then, we obtain
a tridiagonal matrix with diagonal elements $a_i^l$ with $i=1,\ldots,M$
and off-diagonal ones $b_i^l$ with $i=2,\ldots,M$, $a_i^l$ and $b_i^l$
being real.  After diagonalizing the matrix, we obtain ``eigenvalues''
$\epsilon^l_m$ and ``eigenvectors'' $\left|\phi^l_m\right>$
( $m=1,\ldots,M$ ).  Since we stop the Lanczos steps at $M$ steps,
these ``eigenvalues'' and ``eigenvectors'' are approximate ones.
After sampling some arbitrary vectors $\left|\Psi_l\right>$
( $l=1,\ldots, L_0; L_0 \ll L$ ) and repeating the above process
Eq.~(\ref{Lan}), the expectation value is given by
\begin{equation}
\left<\hat{A}\right>\sim\frac{1}{\tilde{Z}}\sum\limits^{L_0}_{l=1}
\sum\limits^M_{m=1}
e^{-\beta \epsilon^l_m} \left<\Psi_l\left|\phi^l_m\right>\right.
\left<\phi^l_m\left|\hat{A}\right|\Psi_l\right>\;,
\label{O2}
\end{equation}
where $\tilde{Z}=\sum\limits^{L_0}_{l=1}\sum\limits^M_{m=1}
e^{-\beta \epsilon^l_m} \left|\left<\Psi_l\left|
\phi^l_m\right.\right>\right|^2$.  If we chose all vectors of
the basis set, i.e., $L_0=L$ and obtained eigenvectors and
eigenvalues exactly, Eq.~(\ref{O2}) should be equivalent to
Eq.~(\ref{O1}).  Jakli\v{c} and Prelov\v{s}ek have shown that,
by using a random state $\left|r\right>=\sum\limits^L_{l=1}
\beta_{rl} \left|\Psi_l\right>$ ($\beta_{rl}$ is randomly distributed)
instead of $\left|\Psi_l\right>$, the expectation value obtained
using Eq.~(\ref{O2}) agrees well with the exact one even if $L_0$
and $M$ are much smaller than $L$.~\cite{Jaklic1}  Similar to
static quantities Eq.~(\ref{O2}), dynamical quantities of the
operator $\hat{A}$ are given by
\begin{eqnarray}
A(\omega)&\sim&\frac{1}{\tilde{Z}}\sum\limits^{L_0}_{l=1}
\sum\limits^M_{m=1} \sum\limits^M_{n=1}
e^{-\beta \epsilon^l_m} \left<\Psi_l\left|\phi^l_m\right>\right.
\left<\phi^l_m\left|\hat{A}^\dagger\right|\tilde{\phi}^l_n\right>
\nonumber \\
&& \times \left<\tilde{\phi}^l_n\left|\hat{A}\right|\Psi_l\right>
\delta(\omega-\epsilon^l_m+\tilde{\epsilon}^l_n)\;,
\label{Od2}
\end{eqnarray}
where $\left|\tilde{\phi}^l_n\right>$ are approximate eigenvectors
with approximate eigenvalues $\tilde{\epsilon}^l_n$ obtained by
the Lanczos procedure Eq.~(\ref{Lan}) starting from the initial
vector $\left|\psi^l_1\right>=\hat{A}\left|\Psi_l\right>$.

\begin{figure}[t]
\epsfxsize=8.0cm
\centerline{\epsffile{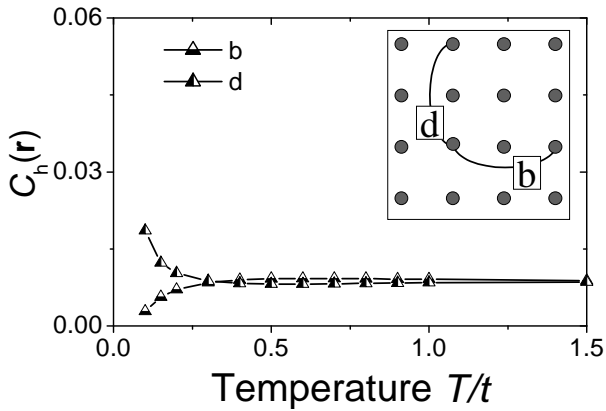}}
\vspace{5mm}
\caption{ Hole correlation function $C_{\rm h}\left(\vec{r}\right)$
in the $t$-$J$ model without the stripe potential on a 4$\times$4
cluster with two holes.  $J/t$=0.4 and $V/t$=0. Labeling configurations
used for the calculation are shown in the inset.}
\label{ChF0}
\end{figure}

In small clusters, there is a characteristic temperature $T^*$ below
which finite-size effects due to the smallness of the systems are
appreciable.  This temperature is approximately proportional to an
average level spacing in the low-energy sector.~\cite{Jaklic2}
In our 4$\times$4 $t$-$J$ cluster without the stripe potential,
the finite-size effects should appear as anisotropic behaviors of
physical quantities along the horizontal and vertical directions
reflecting different boundary conditions along the two directions.
In order to estimate $T^*$, we calculate the hole correlation
function.  Shown in Fig.~\ref{ChF0} is the correlation function
defined as
\begin{eqnarray}
C_{\rm h}\left(\vec{r}\right)= \left<n_i^{\rm h} n_j^{\rm h}
\right>\;,
\label{Ch}
\end{eqnarray}
where $\vec{r}=\vec{R_i}-\vec{R_j}$, $\vec{R_i}$ is the position
vector for the site $i$,  $n_i^{\rm h}=1-n_i$ is the hole-number
operator at site $i$.  In the figure, the hole
correlations with two-lattice spacing labeled {\it b} and {\it d}
are plotted, because the two correlations should be equivalent
in magnitude in two-dimensional systems.  The magnitude of the
correlation in the {\it b} configuration is almost identical to
that of {\it d} above $T$$\simeq$0.3$t$.  However, below
$T$$\simeq$0.3$t$, the two correlations show different temperature
dependence.  This is due to the difference of the boundary
conditions imposed on the cluster.  Therefore the characteristic
temperature $T^*$ in the $t$-$J$ cluster without the stripe
potential is about 0.3$t$.  Even in the presence of the stripe
potential, we assume that $T^*$ does not change so much from
0.3$t$.  In fact, the average level spacing in the low-energy
sector is similar.  In the following, we will show the results
for $T/t$$\geq$0.3.

\begin{figure}[t]
\epsfxsize=8.0cm
\centerline{\epsffile{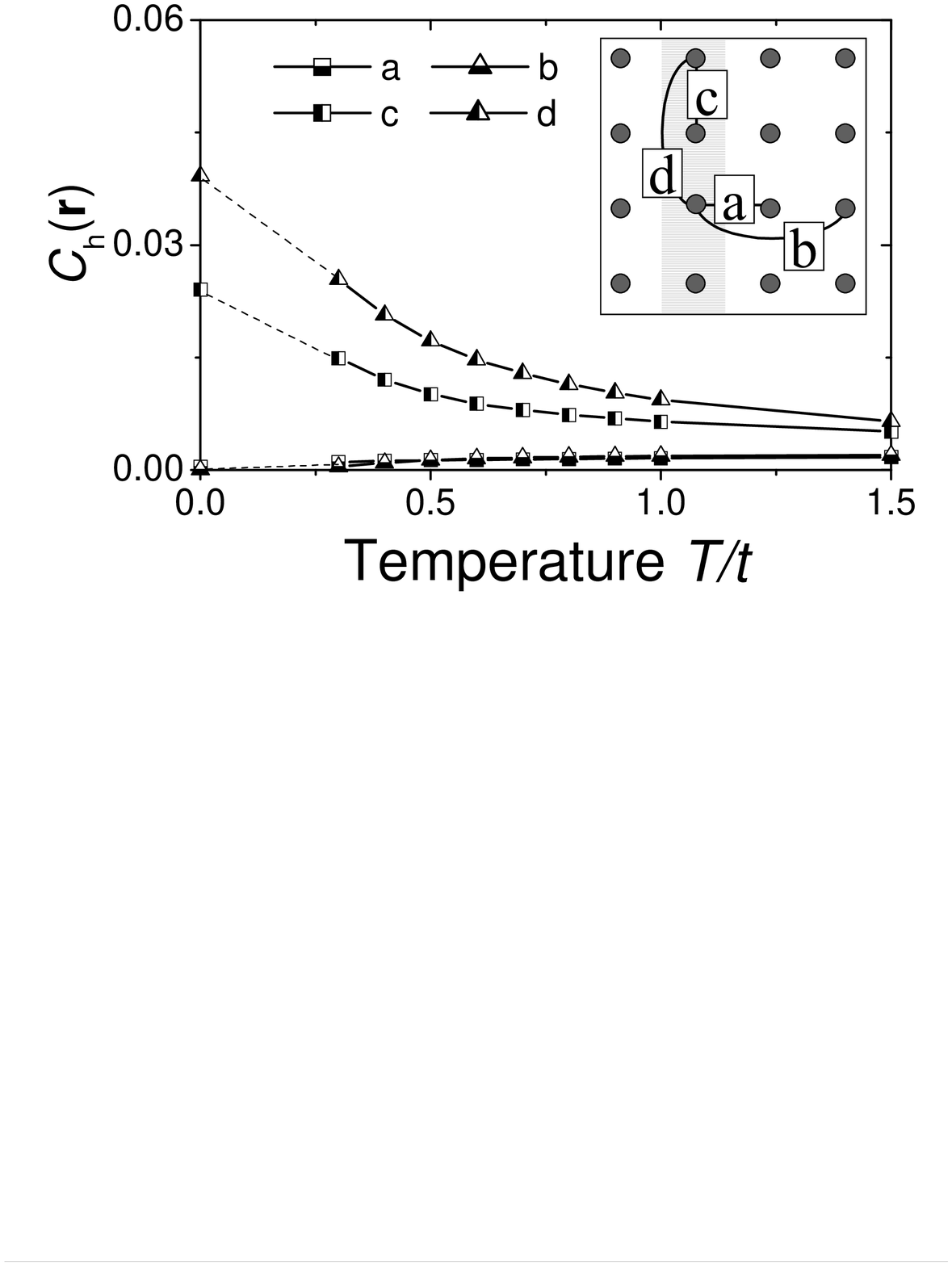}}
\vspace{5mm}
\caption{ Hole correlation function $C_{\rm h}\left(\vec{r}\right)$
in the $t$-$J$ model with the stripe potential on a 4$\times$4
cluster with two holes.  $J/t$=0.4 and $V/t$=1.0. Labeling
configurations used for the calculation are shown in the inset.
The shadow in the inset denotes the direction of the vertical
charge stripe. All dotted lines are guide to eye.}
\label{ChF}
\end{figure}

At $T=0$, we employ the standard Lanczos technique to calculate
various quantities.  In this case, it is necessary to examine
the dependence of given quantities on size and boundary condition
in order to estimate finite-size effects.  We have examined
the dependence by using other clusters with two holes (5$\times$4
with the same boundary condition and 4$\times$4 with periodic
boundary condition along both the directions) and have found
that the hole and spin-correlation functions discussed in
the next section depend only weakly on cluster size and
boundary condition.

\section{Results and Discussion}

Figure~\ref{ChF} shows the temperature dependence of the hole
correlations given by Eq.~(\ref{Ch}) in the $t$-$J$ model with
the stripe potential of $V$/$t$=1.  At $T/t$=0, the correlations
of holes in the vertical directions labeled {\it c} and {\it d}
are larger than those in the horizontal direction labeled {\it a}
and {\it b}.  This anisotropic behavior is due to inhomogeneous
distribution of holes induced by the stripe potential.
With increasing temperature, the correlations in the {\it c} and
{\it d} configurations decrease, while those in {\it a} and
{\it b} increase.  Therefore the anisotropic behavior of the
correlations indicating the confinement of holes in the stripe
becomes less pronounced at finite temperatures.  The evolution
of the hole confinement in the stripe is also observed by
examining the hole occupation number as a function of temperature
(not shown here); the hole number inside (outside) the stripe
decreases (increases) monotonically with increasing temperature.

\begin{figure}[t]
\epsfxsize=8.0cm
\centerline{\epsffile{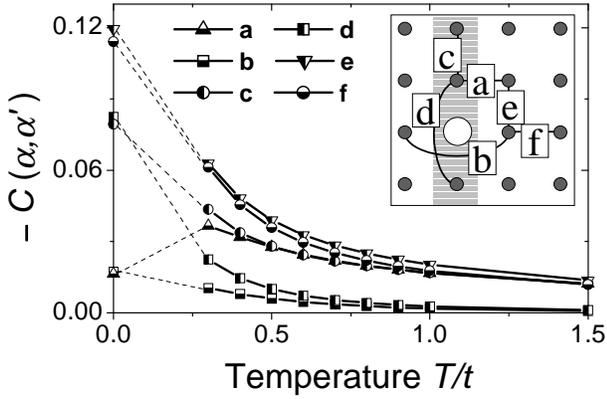}}
\vspace{5mm}
\caption{ Spin-correlation function around a hole
$C\left(\alpha,\alpha^{\prime}\right)$ in the $t$-$J$ model
with the stripe potential on a 4$\times$4 cluster with two holes.
$J/t$=0.4 and $V/t$=1.0.  The open circle in the inset denotes
a hole.}
\label{CF}
\end{figure}

Figure~\ref{CF} shows the results of the spin-correlation function
around a hole,\cite{Tohyama} which is defined as
\begin{equation}
C\left(\alpha,\alpha^\prime\right)
=\frac{1}{N_{\rm h}}\sum\limits_i\left< n_i^{\rm h} S^z_{i+\alpha}
S^z_{i+\alpha^{\prime}}\right>\;,
\label{C}
\end{equation}
where $\alpha$ and $\alpha^{\prime}$ denote two sites around a hole
at site $i$ in the stripe following the labeling convention of
configurations shown in the inset of Fig.~\ref{CF}, and $N_{\rm h}$
is the number of holes ($N_{\rm h}$=2).  At high temperatures,
all of the nearest-neighbor spin correlations in the {\it a},
{\it c}, {\it e}, and {\it f} configurations are antiferromagnetic
and have the same magnitude, because all spins are disordered
by the thermal fluctuation.  The spin correlations between third
nearest neighbors (the {\it b} and {\it d} configurations) are
the same.  With decreasing temperature, holes begin to be confined
into the charge stripes and the magnitude of the spin correlations
increase.  The charge stripes make spin correlation anisotropic
below $T/t\simeq$0.8.  The temperature dependence of the spin
correlations in the spin domain (the {\it e} and {\it f}
configurations) differs from that of the spin correlations
related with a spin in the charge stripe ({\it a} and {\it c}).
At the same time, the spin correlations in the {\it b} and
{\it d} configurations become different.  In the low-temperature
region ($T/t\leq$0.4), the nearest-neighbor spin correlations
inside the spin domains ({\it e} and {\it f}) show temperature
dependence similar to that in the Heisenberg spin system
(not shown here) because of the confinement of holes in the
charge stripe.

The correlation between a spin in the charge stripe and that
in the spin domain (the {\it a} configuration) in Fig.~\ref{CF}
shows interesting temperature dependence below $T/t\simeq$0.4,
when the lowest-temperature data are connected to the
zero-temperature ones:  The correlation is much suppressed
at $T/t$=0, while other spin correlations ({\it b} to {\it f})
are enhanced. This anomalous temperature dependence of the
{\it a} configuration can be understood in the following way:
At $T/t$=0, holes move along the charge stripe coherently and
large Drude weight is obtained parallel to the
stripe.~\cite{Tohyama}  Spin correlation in the configuration
{\it a} is thus suppressed in order to stabilize the coherent
motion of holes.  With increasing temperature, hole carriers
enter into the spin domain, and thermally fluctuating spins
in the stripe disturb the motion of holes along the stripe.
These effects recover interaction between the stripe and the
spin domain, and thus the spin correlation in the {\it a}
configuration is enhanced with the gain of the exchange energy.
In other words, although the charge stripe and the spin domain
are almost separated at $T/t$=0, the interaction between the
charge stripe and the spin domain is recovered and
two-dimensional electronic system is restored with increasing
temperature. This picture will be confirmed by examining the
optical conductivity as shown below.

\begin{figure}[t]
\epsfxsize=8.0cm
\centerline{\epsffile{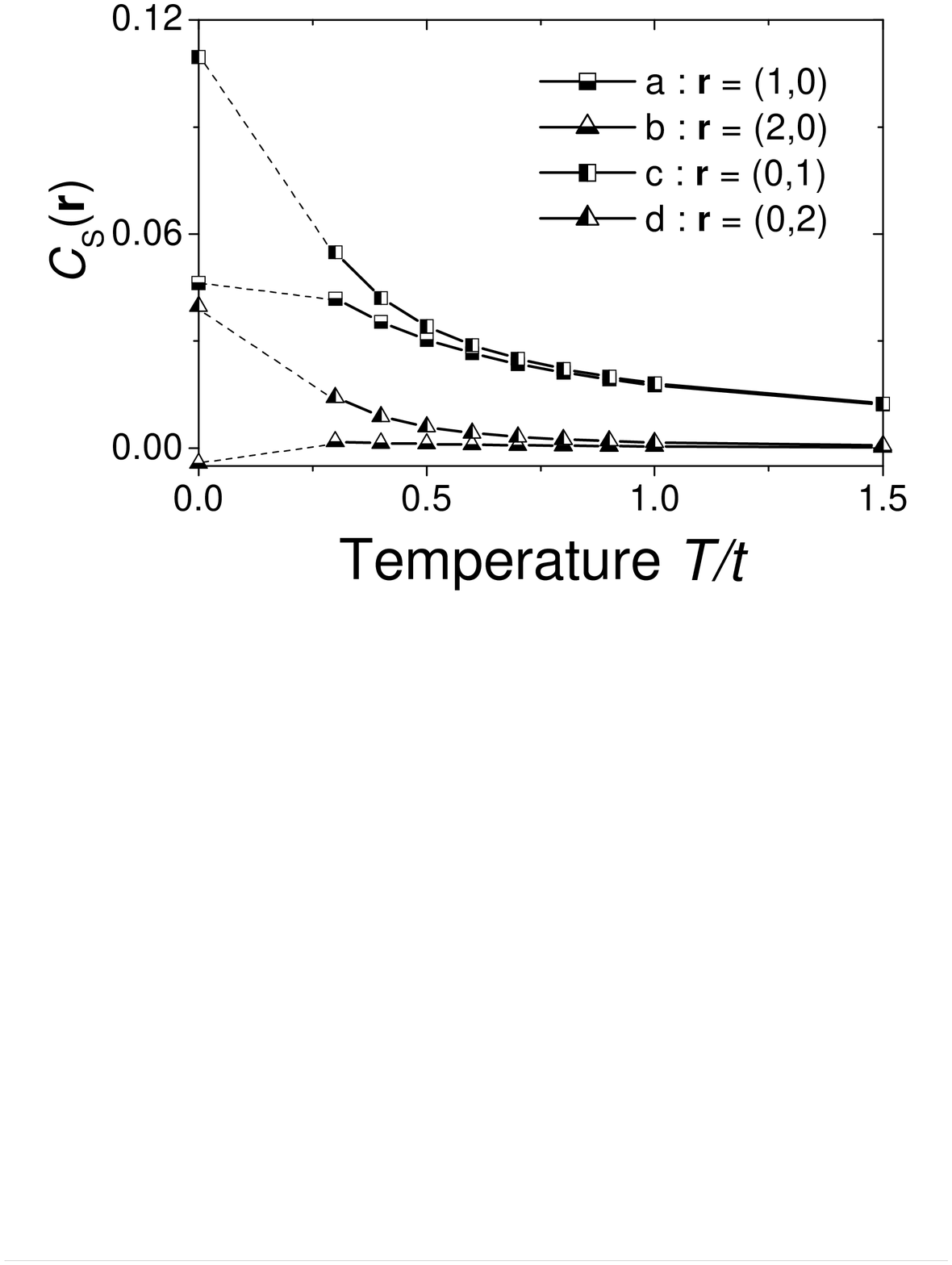}}
\vspace{5mm}
\caption{ Spin-correlation function $C_S\left(\vec{r}\right)$
in the $t$-$J$ model with the stripe potential on a 4$\times$4
cluster with two holes.  $J/t$=0.4 and $V/t$=1.0.}
\label{CsF}
\end{figure}

In Fig.~\ref{CsF}, we show the result of the spin-correlation
function given by
\begin{equation}
C_{\rm S}\left(\vec{r}\right)=
\frac{1}{N_p}\sum\limits_{\left<i,j\right>} \left<{\rm P}\left(i,j\right)
S^z_i S^z_j\right>\;,
\label{Cs}
\end{equation}
where $\vec{r}=\vec{R_i}-\vec{R_j}$, and ${\rm P}\left(i,j\right)$
is 1 when $i$ and $j$ are in the same sublattice and it is $-$1
otherwise.  The summation $\left<i,j\right>$ runs over all the
pairs satisfying a given $\vec{r}$, and $N_p$ is the number of
the pairs.  As seen in the figure, the stripe potential, which
confines holes in the charge stripe, makes the spin correlation
anisotropic.  For example, the spin correlation parallel to the
stripe labeled {\it c} ({\it d}) is larger than that perpendicular
to the stripe labeled {\it a} ({\it b}) below $T/t\simeq$0.8.
With increasing temperature, the anisotropy decreases.

Next, we examine the temperature dependence of the optical
conductivity.  The regular part of the optical conductivity
$\sigma_{\mu\mu}^{\rm reg}\left(\omega>0\right)$
($\mu$=$x$ or $y$) is given by
\begin{eqnarray}
\sigma_{\mu\mu}^{\rm reg}\left(\omega\right)=
\hbar{{1-e^{-\beta\omega}}\over \omega}\Omega_{\mu\mu}\left(\omega\right)
\label{sigma_reg}
\end{eqnarray}
with
\begin{eqnarray}
\Omega_{\mu\mu}\left(\omega\right)=
\frac{\pi}{NZ} \sum\limits_{n\neq m}
e^{-\beta\varepsilon_n}\left|\left<\Phi_n\left|j_\mu\right|\Phi_m
\right>\right|^2 \delta\left(\omega+\varepsilon_m-\varepsilon_n\right)
\label{Omega}
\end{eqnarray}
and
\begin{eqnarray}
j_\mu&=&\frac{iea_0}{\hbar}t\sum\limits_i
\left( \tilde{c}_{i+\mu}^\dagger \tilde{c}_i
-\tilde{c}_i^\dagger \tilde{c}_{i+\mu}\right)\;,
\label{current}
\end{eqnarray}
where $\left|\Phi_n\right>$ is the eigenstate with the eigenvalue
$\varepsilon_n$.  $e$ is the unit of the electric charge and
$a_0$ is the lattice spacing.  Other definitions are standard.
Adding singular contribution to
$\sigma_{\mu\mu}^{\rm reg}\left(\omega\right)$, the real part of
the optical conductivity reads
\begin{eqnarray}
\sigma_{\mu\mu}\left(\omega\right) = 2\pi e^2 D_\mu\delta(\omega)
+ \sigma_{\mu\mu}^{\rm reg} \left(\omega\right)\;,
\label{sigma}
\end{eqnarray}
where the so-called Drude weight $D_\mu$ is given by
\begin{eqnarray}
D_\mu=-\frac{\langle K_\mu\rangle}{2N}-
\frac{1}{\pi e^2} I_{\mu\mu}
\label{Drude}
\end{eqnarray}
with
\begin{eqnarray}
I_{\mu\mu}=\int\nolimits^{\infty}_{+0}\sigma_{\mu\mu}^{\rm reg}
\left(\omega\right)d\omega\;.
\label{sigmainc}
\end{eqnarray}
Here, $\langle K_\mu\rangle$ is the kinetic energy in the $\mu$
direction.  We set $\hbar$=$e$=$a_0$=1 hereafter.
$\Omega_{\mu\mu}(\omega)$ is calculated using Eq.~(\ref{Od2}).

\begin{figure}[t]
\epsfxsize=7.5cm
\centerline{\epsffile{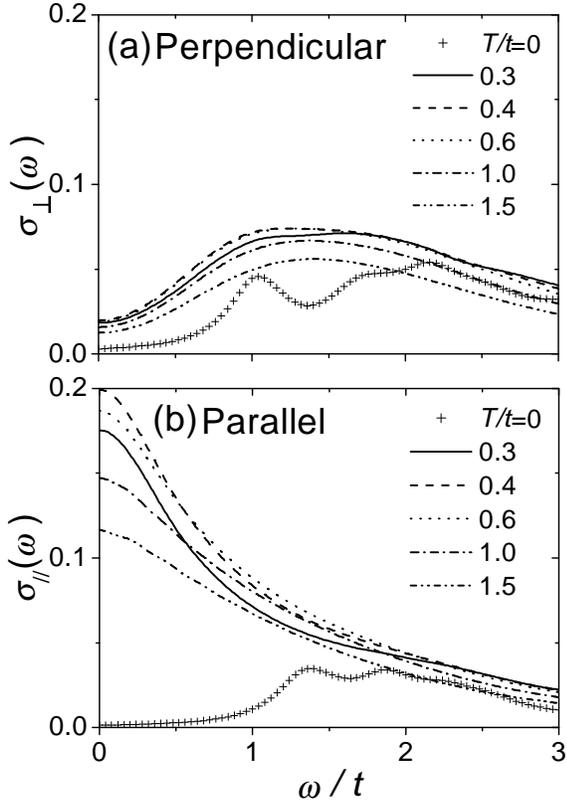}}
\vspace{5mm}
\caption{ Temperature dependence of the regular part of the optical
conductivity in the $t$-$J$ model with the stripe potential on a
4$\times$4 cluster with two holes.  $J/t$=0.4 and $V/t$=1.0.
(a) Perpendicular and (b) parallel to the charge stripe.
Small broadening of 0.2$t$ is used for delta functions.}
\label{OPF1}
\end{figure}

\begin{figure}[t]
\epsfxsize=8.0cm
\centerline{\epsffile{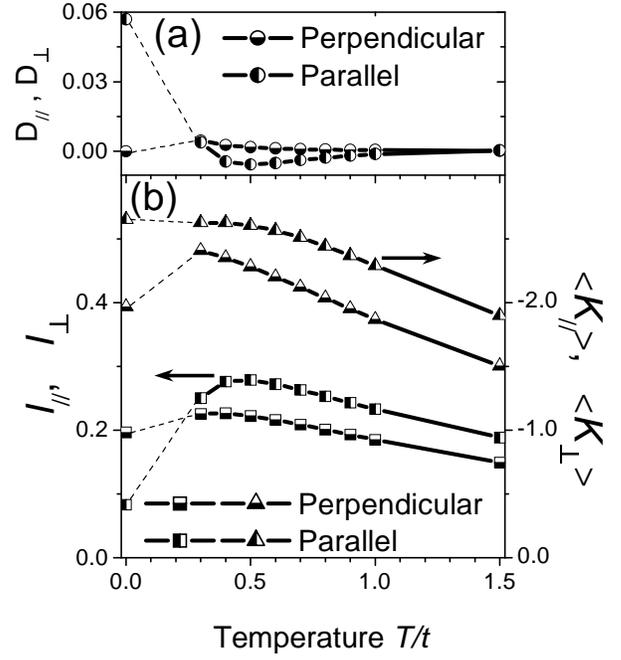}}
\vspace{5mm}
\caption{ (a) Temperature dependence of the Drude weight calculated
by Eq.~(\ref{Drude}) in the $t$-$J$ model with the stripe potential
on a 4$\times$4 cluster with two holes. $J/t$=0.4 and $V/t$=1.0.
(b) Temperature dependence of the integrated spectral weights
$I_\perp$ and $I_{//}$ of the regular part of the optical
conductivity shown in Fig.~\ref{OPF1} (left vertical axis) and
the kinetic energies $\langle K_\perp\rangle$ and
$\langle K_{//}\rangle$ (right vertical axis).
In the text, the temperature at which $I_{//}$ shows a maximum
is denoted by $T_m$.}
\label{OPF2}
\end{figure}

Figures~\ref{OPF1}(a) and (b) show the regular part of the optical
conductivity perpendicular,
$\sigma_\perp^{\rm reg}(\omega)\left[=\sigma_{xx}^{\rm reg}(\omega)\right]$,
and parallel,
$\sigma_{//}^{\rm reg}(\omega)\left[=\sigma_{yy}^{\rm reg}(\omega)\right]$,
to the stripe, respectively. At $T/t$=0, the intensity
of $\sigma_\perp^{\rm reg}(\omega)$ is larger than that of
$\sigma_{//}^{\rm reg}(\omega)$ for $\omega/t\leq$1.0.  In contrast,
in the temperature region of $T/t\geq0.3$, the intensity of
$\sigma_{//}^{\rm reg}(\omega)$ for $\omega/t\leq$1.0 is much
larger than that of $\sigma_\perp^{\rm reg}(\omega)$.
The dramatic increase of $\sigma_{//}^{\rm reg}(\omega)$ from
$T/t$=0 to $T/t$=0.3 is caused by the spectral weight transfer
from the Drude weight parallel to the stripe $D_{//}$ as discussed
below.  Another interesting point in Fig.~\ref{OPF1}(b) is that the
intensity of $\sigma_{//}^{\rm reg}(\omega)$ increases with
increasing temperature from $T/t$=0.3 to $T/t$=0.4, and then
decreases with further increase of $T$.  The intensity maximum
at $T/t\simeq$0.4 is related to the maximum of $I_{//}$
[$=I_{yy}$] shown in Fig.~\ref{OPF2}(b).  Note that $I_{//}$
is the integrated spectral weight of the regular part of the
optical conductivity $\sigma_{//}^{\rm reg}(\omega)$
[see Eq.~(\ref{sigmainc})].  Here, let us consider the origin of
this maximum. At $T/t$=0, since the motion of holes in the charge
stripe is coherent, $D_{//}$ is very large while $D_\perp$ is
almost zero as shown in Fig.~\ref{OPF2}(a). On the contrary,
$I_{//}$ is suppressed compared with $I_\perp$.  With increasing
temperature, the motion of holes in the stripe becomes incoherent,
because spins in the charge stripe fluctuate thermally as is
evidenced by the spin correlations in the {\it c} and {\it d}
configurations in Fig.~\ref{CF}. As a result, $D_{//}$ is
strongly suppressed, while $I_{//}$ is enhanced with the
increase of temperature.  The kinetic energy along the charge
stripe $\langle K_{//}\rangle$ [$=\langle K_{y}\rangle$] exhibits
small temperature dependence below $T/t\simeq0.5$.  From the sum
rule Eq.~(\ref{Drude}), this means that almost all of the
intensity of $D_{//}$ transfers to the regular part
$\sigma_{//}^{\rm reg}$.  With further increase of temperature
above $T/t\simeq0.5$, $\langle K_{//}\rangle$ gradually decreases.
Since $D_{//}$ is almost zero in such high temperature, the
decrease of $\langle K_{//}\rangle$ leads to the decrease of
$I_{//}$. Therefore $I_{//}$ shows a maximum at the temperature
$T_m$$\simeq$0.45$t$.  We have examined the dependence of $T_m$
on the parameters $V$ and $J$; $T_m/t$$\simeq$0.4, 0.45, and 0.6
for $J/t$=0.2, 0.4, and 0.8, respectively, keeping $V/t$=1.0,
and $T_m/t$$\simeq$0.4, 0.45, and 0.5 for $V/t$=0.5, 1.0, and 2.0,
respectively, keeping $J/t$=0.4.  $T_m$ is found to be dependent
not only on $V$ but also strongly on $J$.  These data
are consistent with a picture that the
spin degree of freedom plays an important role in the charge
dynamics in the stripe phase.

Moreover, we find an interesting temperature dependence of the
kinetic energy in Fig.~\ref{OPF2}(b).  At $T/t$=0,
$\langle K_{//}\rangle$ is larger than $\langle K_\perp\rangle$
as expected.  With increasing temperature,
$\langle K_{//}\rangle$ gradually decreases.  Such a decrease
is also seen in the $t$-$J$ model without the stripe potential
(not shown here). In contrast, $\langle K_\perp\rangle$ at
$T/t$=0.3 is larger than that at $T/t$=0.  This anomalous
increase of $\langle K_\perp\rangle$ is consistent with the
picture proposed above, i.e., one-dimensional nature of the
electronic states due to the charge stripe is destroyed by
the thermal fluctuation of spins and thus two-dimensional
electronic states are restored.

Finally, we propose a possible method to compare the present
results of the temperature dependence of $\sigma(\omega)$
with experimental ones.  The frequency dependence of effective
carrier number $N_{eff}(\omega)$ is experimentally evaluated
from the optical conductivity by using a relation that
$N_{eff}(\omega)=2mV$/$\left(\pi e^2\right)
\int^\omega_0\sigma(\omega')d\omega'$,
where $m$ is the mass of a free electron and $V$ is the volume
of the unit cell.~\cite{Uchida}  From the theoretical side,
the effective carrier number in high-frequency limit
$\omega\rightarrow\infty$ is proportional to the average of
the kinetic energies $\langle K_\perp\rangle$ and
$\langle K_{//}\rangle$ because of a relation that
$\int^\infty_0\sigma(\omega)d\omega=
-\pi e^2\left(\langle K_\perp\rangle+\langle
K_{//}\rangle\right)$/$(4N)$.
Since the dominant part of the temperature-induced change of
the calculated $\sigma_{//}(\omega)$ and $\sigma_\perp(\omega)$
is concentrated in the region of $\omega\lesssim$3$t$
($\simeq$1~eV), the change of the averaged kinetic energies
can be a measure of the change of the effective carrier
number up to around 1~eV.  In a realistic temperature region
($T<J$=0.4$t$), $\langle K_{//}\rangle$ is almost
temperature independent, while $\langle K_\perp\rangle$
increases with increasing $T$ because of the destruction of
the stripe as mentioned above.  Thus the averaged kinetic
energy increases.  We propose that $N_{eff}(\omega)$ at around
$\omega$$\simeq$1~eV in the stripe phase of the high-$T_c$
cuprates increases with increasing temperature.  This may be
detectable from the detailed analysis of the temperature
dependence of experimental data.

\section{SUMMARY}
In summary, we have examined the temperature dependence of the
electronic states with vertical charge stripes. The
spin-correlation function, the optical conductivity, and the kinetic
energy have been calculated by using the finite temperature
Lanczos method.  We have found that the motion of holes along
the charge stripes, which is coherent at $T/t$=0, becomes
incoherent with increasing temperature.  We have also found that
the spin correlation between a spin in the charge stripe and
that in the spin domain is smaller than that for other spin pairs
at $T/t$=0.  However, the correlation increases with increasing
temperature and then decreases toward high temperature.  This
anomalous increase of the correlation is a manifestation of
the evolution to the two-dimensional electronic states.
Moreover, we have found an anomalous change of the kinetic
energy perpendicular to the stripe.  All of these characteristic
phenomena of the stripe phase appear in a realistic temperature
region, i.e., $T<J$.  The fact that characteristic temperatures
such as $T_m$ defined in Fig.~\ref{OPF2}(b) depend not only on
the stripe potential $V$ but also on $J$ suggests that the
ordering of spins plays an important role in the confinement of
holes in the charge stripe.  The disordering of spins due to
the thermal fluctuation destroys the charge stripe and makes
the crossover from the one-dimensional electronic states
to the two-dimensional ones.

\acknowledgements
We would like to thank P. Prelov\v{s}ek for discussion and
for giving us the information of the finite temperature
Lanczos method.  This work was supported by a Grant-in-Aid for
Scientific Research on Priority Areas from the Ministry of
Education, Culture, Sports, Science and Technology of Japan,
CREST, and NEDO.  The numerical calculations were performed
in the Supercomputer Center, Institute for Solid State Physics,
University of Tokyo, and the supercomputing facilities in
Institute for Materials Research, Tohoku University.


\begin{references}
\bibitem{AntiPhase}
We also examined a $4\times 4$ cluster of the $t$-$J$ model with
periodic boundary condition both parallel and perpendicular to the
charge stripe.  Although the periodic boundary condition perpendicular
to the charge stripe causes frustration of spins, our conclusions are
not affected by the boundary condition.
\end{references}
\end{document}